\documentclass[11pt,A4,twoside]{article}
\usepackage[latin1]{inputenc}
\usepackage{amsmath}
\usepackage{bbm}
\usepackage{natbib}
\usepackage{eucal}
\usepackage{graphicx}
\usepackage{titlesec}
\usepackage[english]{babel}
\usepackage{times}
\usepackage{titling}
\usepackage{float}
\usepackage[font=small]{caption}
\usepackage{array}
\usepackage[colorlinks,citecolor=blue,urlcolor=blue]{hyperref}
\usepackage{authblk}
\usepackage{xcolor}

\usepackage{boldline}

\usepackage{setspace}
\DeclareSymbolFont{txgreek}{OML}{cmr}{m}{it}

\renewcommand{\abstract}[1]{{\small\noindent
\hrulefill\par \vspace*{0.1cm}\noindent{\small\bf\sffamily
{Abstract}}\parindent=0pt\par\noindent\vspace{-0.1cm}\noindent\hrulefill\par\vspace*{0.5\baselineskip}\hspace*{0cm}\renewcommand{\baselinestretch}{1.1}\sffamily{#1}\par\vspace*{-0.1cm}\noindent\hrulefill}}

\def\and{,\;}
\DeclareMathSizes{11}{11}{7.8}{6.5}

\def\paragraf{\fontsize{9}{10pt}\fontfamily{phv}\fontshape{it}\selectfont}
\titleformat{\paragraph}
{\titlerule[0pt]
\vspace{0cm}%
\paragraf}
{\theparagraph}{.5em}{\vspace*{0\baselineskip}}

\def\titol{\fontsize{12.045}{12pt}\fontfamily{phv}\fontseries{b}\selectfont}
\titleformat{\section}
{\titlerule[0pt]
\vspace{0.2cm}%
\titol}
{\thesection.}{.5em}{\vspace*{0\baselineskip}}

\def\titolp{\fontsize{11.045}{11pt}\fontfamily{phv}\fontseries{b}\fontshape{it}\selectfont}
\titleformat{\subsection}
{\titlerule[0pt]\bigskip
\vspace{-0.4cm}%
\titolp}
{\thesubsection.}{.5em}{\vspace*{0\baselineskip}}

\def\titolpp{\fontsize{10.045}{10pt}\fontfamily{phv}\fontshape{it}\selectfont}
\titleformat{\subsubsection}
{\titlerule[0pt]
\vspace{-0.4cm}%
\titolpp}
{\thesubsubsection.}{.5em}{\vspace*{0\baselineskip}}

\usepackage{titling}
    \pretitle{\begin{center}\sffamily\fontsize{18pt}{20pt}\selectfont}
    \posttitle{\par\end{center}}
    \preauthor{\begin{center}\fontsize{12pt}{14pt}\selectfont}
    \postauthor{\par\end{center}\vspace{0bp}}
    \predate{}
    \date{}
    \postdate{}

\title{Compositional Covariance Shrinkage and Regularised Partial Correlations}

\thanksmarkseries{arabic}
\author{Suzanne Jin\thanks{Centre for Genomic Regulation (CRG), Dr Aiguader, 88, 08003 Barcelona, Spain}$^{~,2}$ \and  C\'edric Notredame$^{1,}$\thanks{Universitat Pompeu Fabra, Barcelona, Spain} \and Ionas Erb$^{1,}$\thanks{Corresponding author: \texttt{ionas.erb@crg.eu}}}

\def\headers#1{\fontsize{8.5}{10}\centering\sffamily\itshape{#1}}
\def\page#1{\fontsize{8.5}{10}\sffamily{#1}}
\usepackage{fancyhdr}

\begin{document}
\maketitle

\thispagestyle{empty}
\renewcommand{\headrulewidth}{0truecm}
\pagestyle{fancy}
\rhead[\headers{Compositional Covariance Shrinkage}]{\page{\thepage}}
\lhead[\page{\thepage}]{\headers{Jin et al.}}
 \lfoot{} \rfoot{}
\cfoot{}

\abstract{We propose an estimation procedure for covariation in wide compositional data sets. For compositions, widely-used logratio variables are interdependent due to a common reference. \emph{Logratio uncorrelated} compositions are linearly independent before the unit-sum constraint is imposed. We show how they are used to construct bespoke shrinkage targets for logratio covariance matrices and test a simple procedure for partial correlation estimates on both a simulated and a single-cell gene expression data set. For the underlying counts, different zero imputations are evaluated. The partial correlation induced by the closure is derived analytically. {Data and code are available from GitHub.}}

\paragraph{MSC: G2F30, G2H20, G2P99}

\paragraph{Keywords: Compositional Covariance Structure, Logratio Analysis, Partial Correlation, James-Stein Shrinkage}

\renewcommand{\baselinestretch}{1.2}
\bigskip

\section{{Introduction}}

The study of data variability is at the heart of data analysis. A simple way of quantifying the variability around some central tendency is given by a bilinear function defined for each pair of variables: the covariance. The distribution that makes no assumptions beyond a finite covariance matrix is the multivariate Gaussian, and it is thus optimal in a maximum-entropy sense. Covariance matrices can capture the relevant aspects of variability in a wide variety of scenarios \cite{Whittaker}, and the Gaussian doesn't have to be the ``true" underlying distribution for this\footnote{For a discussion covering this aspect of maximum-entropy approaches see \cite{vanNimwegen}. A more general maximum-entropy approach to compositional data has recently been proposed \cite{maxEnt}, but it is difficult to obtain analytic results from it.}.\\ 
Often the limitations of data acquisition make good covariance estimates difficult, and regularisation techniques are adopted to deal with small data sets. One such technique is covariance \emph{shrinkage}, which is a regularisation technique that lets us invert singular covariance matrices which arise from the presence of fewer samples than variables \cite{SchaeferStrimmer}. It has recently been applied in the context of \emph{compositional} data in genomics \cite{Badri}{, where clear improvements over empirical measures were shown with a comprehensive and reproducible benchmark.} Compositional data are constrained in that their samples sum to a constant, which causes biases that can be alleviated by logratio transformations \cite{AitchisonBook}. \cite{Badri} {applied the standard shrinkage approach directly to logratio covariance matrices, which does not take into account the interdependence among logratio variables.} Our main interest in this contribution is to provide and test an improved shrinkage procedure that is {custom-made} for logratio-transformed data. The properly regularized logratio covariance matrices can then be inverted to estimate logratio-based partial correlations, which we have proposed as a measure of interaction between compositional variables \cite{Erb2020}.\\
{On a more general note, we aim to contribute to finding a principled way of quantifying interactions in data sets consisting of compositions or relative counts. When inferring interactions between variables, a useful distinction (see, e.g., \cite{Werhli}) is made between relevance networks, which are based on pairwise interaction coefficients (such as Pearson correlation) and graphical models, which make use of all the variables to infer pairwise association (e.g., via partial correlation). While the former are susceptible to inferring indirect interactions that are mediated by variables that are unrelated to the pair of variables in question, the latter aim to infer the conditional independence structure of their joint distribution \cite{Whittaker}. This way they can in principle infer direct interactions between two variables. The distinction is highly relevant for compositional data, where variables are intrinsically connected due to the unit-sum constraint. Here, simple relevance networks are inadequate because of the negative bias coming from the constraint. This bias has been addressed by measures that use ratios of variables, such as Aitchison's logratio variance \cite{AitchisonBook}, as well as other coefficients inferring proportionality \cite{Lovell2015,ErbNotredame}. These can be used to infer relevance networks in compositional data, and \cite{Badri} have shown how shrinkage improves estimates of the proportionality coefficient. However, to define graphical models for compositions similar to \cite{Kurtz}, logratio-based partial correlations will be instrumental.}\\
{In what follows, in section 2 we briefly review how the covariance structure and partial correlations are described in the logratio framework and outline James-Stein shrinkage of general covariance matrices as proposed in \cite{SchaeferStrimmer}. In Section 3, the concept of logratio uncorrelated compositions introduced by Aitchison is described, and it is shown how one can make use of it to define shrinkage for logratio covariance matrices. Two equivalent approaches are proposed: 1) shrinking a so-called \emph{basis} covariance matrix which is then transformed into a logratio covariance, and 2) shrinking logratio covariance matrices directly with the help of logratio uncorrelated targets. Due to its simplicity, we prefer approach 1, and both the synthetic and the experimental data benchmark of Section 4 will be based on it. In section 5, we discuss how the use of partial correlations relates to normalisation techniques in genomics and provide an expression for the remaining partial correlation in logratio-uncorrelated compositions. A summary section and an Appendix containing some of the more technical details conclude the paper.}     

\section{Preliminaries}
\subsection{Compositional Data and Logratio Covariance}
Let us denote $N$ compositional samples by vectors $\boldsymbol{p}_i$, $i=1,\dots,N$, whose $D$ positive components (called {\it parts}) sum to 1. Specifically, we consider data with a large number of parts $D$ compared with the number of samples $N$. Such wide compositional data are of interest in genomics \cite{specialIssue}, where they appear mainly in the form of relative counts coming from sequencing experiments \cite{OutlookReview}. In this case we would consider $\boldsymbol{p}$ the closed data obtained when dividing a count sample by its total, i.e., the compositions are the empirical parameter estimates of a multinomial model.\\
The counts play the role of a so-called \emph{basis} of the composition, a concept that was introduced in \cite{Aitchison1982}, and which we will talk about in more detail in section \ref{LUsection}. Considering the counts as \emph{relative} data is justified by assigning no importance to the variations in the count totals (which correspond to the \emph{size} of the basis). Compared with real numbers, relative counts constitute a \emph{discrete} basis, which implies some problems of its own, especially when counts are small \cite{LovellNARGAB} or vanish entirely. Such problems can be alleviated with suitable weights \cite{voom} or with finite-size corrections to the frequencies that can again be obtained via shrinkage \cite{HausserStrimmer}. The latter, known as frequency shrinkage, has the advantage that it imputes count zeros naturally. The promise that this has shown for compositional data \cite{Amalgams,Reappraisal} will be confirmed here in our benchmarks.\\
Let us now consider an $N\times (D-1)$ data matrix $\boldsymbol{X}$ with real-valued elements. These elements are obtained from the log-ratios we evaluate from our compositional samples $\boldsymbol{p}_i$ using the additive logratio transformation (ALR, introduced in \cite{Aitchison1982})
\begin{equation}
    x_{ij}=\log\frac{p_{ij}}{p_{iD}},\qquad j=1,\dots,D-1.\label{alr}
\end{equation}
The (unbiased) sample covariance matrix $\boldsymbol{S}$ is then defined by its elements 
\begin{equation}
    s_{ij}=\frac{1}{N-1}\sum_{k=1}^N(x_{ki}-\bar{x}_i)(x_{kj}-\bar{x}_j),\label{S}
\end{equation}
where $\bar{x}_i$ are the components of the $N$-dimensional vector $\bar{\boldsymbol{x}}$ of column means of $\boldsymbol{X}$. In order to have a criterion allowing us to judge the quality of our covariance estimate based on $\boldsymbol{S}$, we need a distributional assumption. Let us assume that $\boldsymbol{X}$ was sampled from  a normal distribution with population parameters $\boldsymbol{\mu}$ and $\boldsymbol{\Sigma}$. This can be written as
\begin{equation}
    \boldsymbol{x}\sim\mathcal{N}(\boldsymbol{\mu},\boldsymbol{\Sigma}),\label{Gauss}
\end{equation}
where $\boldsymbol{x}$ denotes a $D-1$-dimensional random vector taking values as in the rows of $\boldsymbol{X}$. In this case the random compositions $\boldsymbol{p}$ are distributed according to Aitchison's celebrated logistic normal distribution \cite{Aitchison1982}. If we denote the density of the compositions themselves by $f_\mathcal{L}$, the Gaussian density $f_\mathcal{N}$ will be multiplied by an additional factor due to the resulting Jacobian:
\begin{equation}
    f_\mathcal{L}(\boldsymbol{p}=\boldsymbol{p}_i|\boldsymbol{\mu},\boldsymbol{\Sigma})=\left(\prod_{j=1}^Dp_{ij}\right)^{-1}f_\mathcal{N}(\boldsymbol{x}=\boldsymbol{x}_i|\boldsymbol{\mu},\boldsymbol{\Sigma}),
\end{equation}
where $\boldsymbol{x}_i$ is a function of $\boldsymbol{p}_i$ defined by (\ref{alr}).

\subsection{Covariance Shrinkage}

To improve estimates of $\boldsymbol{\Sigma}$ from data, {adding a bit of bias to the} unbiased estimator $\boldsymbol{S}$ can {reduce its overall error. This is a consequence of the bias-variance decomposition of the mean-squared error and can be achieved by} convexly combining $\boldsymbol{S}$ with (a.k.a.\ ``shrinking it towards") a suitable target matrix $\boldsymbol{T}$:
\begin{equation}
    \hat{\boldsymbol{\Sigma}}=\lambda\boldsymbol{T}+(1-\lambda)\boldsymbol{S},\label{shrink}
\end{equation}
where the shrinkage intensity $\lambda$ is between zero and one. $\hat{\boldsymbol{\Sigma}}$ is a so-called James-Stein type estimator \cite{JamesStein}. If we forget for a moment that $\boldsymbol{X}$ was obtained from logratios of compositions, a typical target matrix could be $\mathrm{diag}(\boldsymbol{S})$.  While (\ref{shrink}) tells us nothing about suitable values of $\lambda$, its optimal value can be estimated from the data via an analytic expression that minimises a mean-squared error cost-function with respect to the population covariance \cite{LedoitWolf}. The derivation of this expression makes use of the fact that $\boldsymbol{S}$ is unbiased. This optimum is given by
\begin{equation}
    \lambda^*=\frac{\sum_{j=1}^D\sum_{i\ne j}\left[\mathrm{var}(s_{ij})-\mathrm{cov}(s_{ij},\tau_{ij})\right]}{\sum_{j=1}^D\sum_{i\ne j}\mathbbm{E}\left[(s_{ij}-\tau_{ij})^2\right]},\label{lambda}
\end{equation}
where $\tau_{ij}$ denotes the matrix entries of $\boldsymbol{T}$. {Note that here, covariance and expectation of the matrix elements refer to parameters of a distribution which can be approximated by their empirical estimates for actual calculations}, see \cite{SchaeferStrimmer}.\\
To provide further background, we also show a principled interpretation of covariance shrinkage in the Appendix. There, we sketch how covariance shrinkage is equivalent to optimising the posterior probability of $\boldsymbol{\Sigma}$ from data and prior information. Note that $\lambda^*$ is inferred from data, making this a quasi-empirical Bayes approach.\\
{The shrunk covariance estimator is useful in itself as it often by far outperforms the sample covariance unless the number of samples considerably exceeds the number of variables. Regularisation becomes a necessity for inverting the covariance matrix whenever the number of samples is smaller than the number of variables. Therefore, shrinkage improves and extends the range of applications of the analysis proposed in the next section.} 

\subsection{Logratio-based Partial Correlations}

Partial correlations between two variables are evaluated by controlling for the linear dependence on all the remaining variables, e.g., \cite{Whittaker}. This is achieved by correlating the residuals of two variables, where the residuals are with respect to the linear least squares predictors obtained from the remaining variables. Intuitively, we only correlate those contributions to our variables that are orthogonal to a subspace spanned by the variables we control for. Partial correlations can therefore be used to extract direct dependencies between variables, as opposed to the ones based on Pearson correlation, which can be dominated by indirect interactions via other variables. Partial correlations between {\it logratio} variables can be defined independently of their reference part $D$, so they measure association directly between parts, provided their reference is controlled for \cite{Erb2020}. Let us recall the formula for the (logratio-based) partial correlation between two parts $i,j$ of the random composition $\boldsymbol{p}$ via their ALR covariance $\boldsymbol{\Sigma}$. {Below, the backslash means taking the set difference, so the variables we control for are all the ones not belonging to the pair in question:}
\begin{equation}
r_{ij}(\boldsymbol{p}):=\mathrm{corr}(x_i,x_j|\left\{x_1,\dots,x_{D-1}\right\}\backslash\{x_i,x_j\})=\frac{-\sigma^{(-1)}_{ij}}{\sqrt{\sigma^{(-1)}_{ii}\sigma^{(-1)}_{jj}}},\label{partcorr}
\end{equation}
where $i\ne j$ and $i,j=1,\dots,D-1$, and where $\sigma^{(-1)}_{ij}$ denotes the elements of the inverse of $\boldsymbol{\Sigma}$. It may appear as if the reference part $p_D$ is ``sacrificed" here, and its partial correlations cannot be determined. Note, however, that $r_{ij}$ is invariant under permutation of the reference part: all $D$ choices of $\boldsymbol{\Sigma}$ lead to the same partial correlations (for the parts they have in common). It follows that we can obtain the partial correlation of any pair of parts by choosing an arbitrary reference part that doesn't form part of the pair. {We have stated our definition in terms of  the ALR covariance matrix because it comes closest to what is known from unconstrained data. There is also a {(computationally)} more convenient representation using a covariance matrix involving all parts in form of centred logratios (the CLR covariance introduced in section \ref{LUsection})}, from which the partial correlations can be obtained via its pseudoinverse. The inverse can also be estimated with a sparsity assumption, which has been done to create ecological networks of microbes, see \cite{Kurtz}. \enlargethispage{\baselineskip}

\subsection{Linear Independence in Logratio and Basis Covariance Matrices}\label{LUsection}

\begin{figure}[H]
\begin{center}
\includegraphics[width=0.6\textwidth]{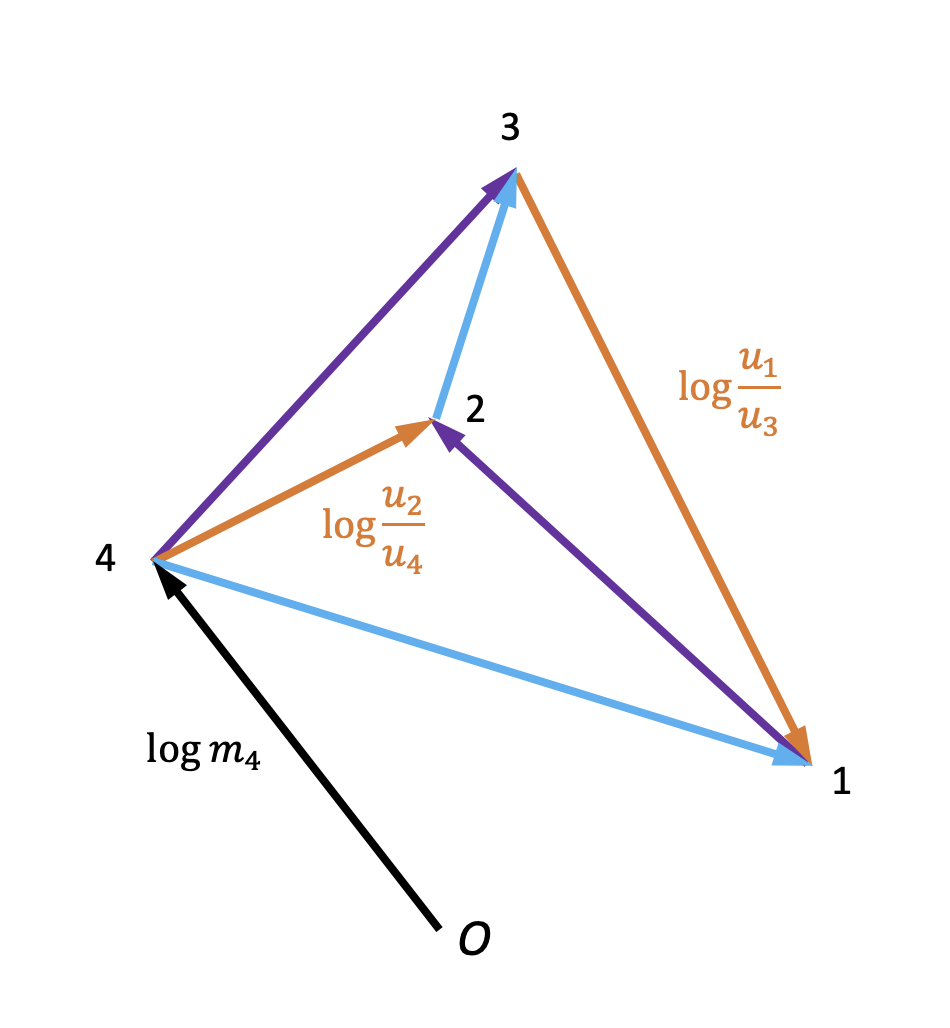}
\caption{Dependence structure in a logratio uncorrelated composition $\boldsymbol{u}$ with four parts. Transformed parts can be visualized as centred vectors { with sampled components (labels are in random-variable notation)}, where length corresponds to standard deviation. The reference part is indexed by $D=4$ and  its basis {$m_4=tu_4$} is usually unknown (resulting in an unknown origin $O$). The vectors from the origin to the points indexed by 1 to 4 are the log-transformed basis vectors, their squared lengths correspond to the $\alpha_k$ defined in (\ref{alpha}). As $\boldsymbol{u}$ is LU, they are orthogonal to each other in four dimensions. In three dimensions, only the logratio vectors of equal colour are orthogonal (e.g., the orange vectors labelled with their logratios). These correspond to vectors where all indices are different, see (\ref{LUcondition}).}\label{LUcomposition}
\end{center}
\end{figure}

For covariance matrices of unconstrained data, vanishing off-diagonal elements mean linear independence of the variables. Logratio covariance matrices cannot be linearly independent in this sense. Although logratio transformations map compositional data from the simplex to a real-valued space, they cannot remove the dependence between the parts that is introduced by the loss of one dimension. The notion coming closest to linear independence for logratios is given by {\it logratio uncorrelated} compositions. In the following, we state some results from \cite{AitchisonBook}, section 5.9, where this concept is introduced. A composition $\boldsymbol{p}$ is said to be logratio uncorrelated (LU) if
\begin{equation}
    \mathrm{cov}\left(\log\frac{p_i}{p_k},\log\frac{p_j}{p_l}\right)=0\label{LUcondition}
\end{equation}
for every selection of \emph{four different} indices $i,j,k,l$ from $\{1,\dots,D\}$. {This case is shown in Figure} \ref{LUcomposition}.
The covariance structure of LU compositions can be made explicit when defining a $D$-component vector $\boldsymbol{\alpha}$ with elements
\begin{equation}
    \alpha_k=\mathrm{cov}\left(\log\frac{p_i}{p_k},\log\frac{p_j}{p_k}\right),\label{alpha}
\end{equation}
with the three indices $i$, $j$, $k$ {in $\{1,\dots,D\}$ and all different from each other}. For LU compositions, this definition makes sense because the value of $\alpha_k$ is constant regardless of the (unequal) indices $i,j$. The matrix $\boldsymbol{\Sigma}$ of an LU composition is now given by
\begin{equation}
    \sigma_{ij} =
    \left\{
      \begin{array}{c@{\quad}l}
         \alpha_i+\alpha_D & \mbox{if $i=j$,} \\
         \alpha_D
         & \mbox{if $i\ne j$.}
      \end{array}
    \right.\qquad i,j = 1,\dots,D-1.\label{LRuncorr}
\end{equation}
(we reproduce the proof of this in the Appendix). While these definitions may seem overly abstract, they appear much more compelling in the light of Aitchison's concept of the \emph{basis} of a composition, see \cite{AitchisonBook}, chapter 9. Intuitively, the basis is one of many ways our compositions can look like before the closure operation is applied. It is defined by $\boldsymbol{m}=t\boldsymbol{p}$ for some positive real $t$ which is called the {\it size} of the basis and {for which} $t=\sum_{k=1}^D m_k$ {holds (implying that it can change for each data sample)}. Of course, infinitely many bases of different size exist for a given unit-sum composition. Now the \emph{basis covariance} matrix $\boldsymbol{\Omega}$ is defined to have the elements
\begin{equation}
    \omega_{ij}=\mathrm{cov}\left(\log m_i,\log m_j\right),
\end{equation}
from which the covariance structure of a composition can be obtained unambiguously. If we assume that $\boldsymbol{\Omega}$ is diagonal, i.e., the logged components of the basis vector are mutually uncorrelated, {we have the result that} $\boldsymbol{\alpha}$ coincides with $\mathrm{var}(\log\boldsymbol{m})$:
\begin{equation}
    \alpha_i=\omega_{ii}, \qquad i=1,\dots,D.\label{alphaOmega}
\end{equation}
This goes to show that $\boldsymbol{\alpha}$ can be seen as the variance vector of an independent distribution in a sample space that has one additional dimension (and in which the logratio sample space is embedded), see Figure \ref{LUcomposition}. As $\boldsymbol{\alpha}$ was defined for logratios, this statement may appear confusing because of the apparent dependence of $\mathrm{var}\log(t\boldsymbol{p})$ on the basis size $t$. However, the $t$-dependence drops out because $\boldsymbol{\Omega}$ is diagonal (see {the proof of (\ref{alphaOmega}) in the} Appendix).\\ 
For completeness, let us also mention an equivalent expression to (\ref{LRuncorr}) in terms of the centred logratio transformation (CLR). Here the reference involves all the parts of the composition. It transforms the compositional data matrix as
\begin{equation}
    y_{ij}=\log\frac{p_{ij}}{g(\boldsymbol{p}_i)},
\end{equation}
where $g(\boldsymbol{p}_i)=\prod_{k=1}^Dp_{ik}^{1/D}$ is the geometric mean of the $i$-th composition. Note that the resulting data matrix $\boldsymbol{Y}$ has a constraint $\sum_{k=1}^Dy_{ik}=0$ acting on its rows. This constraint leads to a singular CLR covariance matrix of rank $D-1$. The covariance of $\boldsymbol{y}$ is denoted by $\boldsymbol{\Gamma}$. The two forms of log-ratio covariance are equivalent and can be transformed into each other (see the first two rows in Table \ref{ttable}).
Using the transformation from $\boldsymbol{\Sigma}$ to $\boldsymbol{\Gamma}$ on (\ref{LRuncorr}), we obtain the elements of $\boldsymbol{\Gamma}$ for an LU composition as
\begin{equation}
    \gamma_{ij}=\left\{
      \begin{array}{c@{\quad}l}
         \alpha_i-(2\alpha_i-\frac{1}{D}\sum_k\alpha_k)/D & \mbox{if $i=j$,} \\
         -(\alpha_i+\alpha_j-\frac{1}{D}\sum_k\alpha_k)/D
         & \mbox{if $i\ne j$.}
      \end{array}
    \right.\qquad i,j = 1,\dots,D.\label{GammaLU}
\end{equation}

\begin{table}[h]
\centering
    \begin{tabular}{|p{1.5cm}| p{9.2cm}|}
 \hline
$\boldsymbol{\Sigma}\rightarrow\boldsymbol{\Gamma}$ &  $\gamma_{ij}=\sigma_{ij}-\frac{1}{D}\sum_{k=1}^D\sigma_{ik}-\frac{1}{D}\sum_{k=1}^D\sigma_{kj}+\frac{1}{D^2}\sum_{k,l=1}^D\sigma_{kl}$ \\
 \hline
$\boldsymbol{\Gamma}\rightarrow\boldsymbol{\Sigma}$ & $\sigma_{ij}=\gamma_{ij}-\gamma_{iD}-\gamma_{Dj}+\gamma_{DD}$  \\
\hlineB{3.5}
$\boldsymbol{\Omega}\rightarrow\boldsymbol{\Sigma}$ & $\sigma_{ij}=\omega_{ij}-\omega_{iD}-\omega_{Dj}+\omega_{DD}$\\
\hline
$\boldsymbol{\Sigma}\rightarrow\boldsymbol{\Omega}$ & $\omega_{ij}=\gamma_{ij}(\boldsymbol{\Sigma})+\beta_i+\beta_j$,\qquad where\\
& $\beta_j=\mathrm{cov}\left(\mathrm{clr}_j(\boldsymbol{p}),\log g(\boldsymbol{m})\right)-\frac{1}{2}\mathrm{var}\left(\log g(\boldsymbol{m})\right)$.\\
\hline
$\boldsymbol{\Omega}\rightarrow\boldsymbol{\Gamma}$ & $\gamma_{ij}=\omega_{ij}-\frac{1}{D}\sum_{k=1}^D\omega_{ik}-\frac{1}{D}\sum_{k=1}^D\omega_{kj}+\frac{1}{D^2}\sum_{k,l=1}^D\omega_{kl}$\\
\hline
$\boldsymbol{\Gamma}\rightarrow\boldsymbol{\Omega}$ & $\omega_{ij}=\gamma_{ij}+\beta_i+\beta_j$,\qquad see $\boldsymbol{\Sigma}\rightarrow\boldsymbol{\Omega}$.\\
\hline
\end{tabular}
\caption{Elementwise transformations between ALR, CLR, and basis covariance matrices (collected from \cite{AitchisonBook}). Here, indices of $\sigma_{ij}$ are allowed to take the value $D$, in which case $\sigma_{ij}$ vanishes; clr$_j$ stands for the $j$-th element of the CLR transform, and $g$ denotes the geometric mean. All these transformations can also be expressed as simple matrix operations that are not shown here.}
\label{ttable}
\end{table}

\section{Logratio Covariance Shrinkage}\label{LRshrinkage}

The specific covariance structure of compositional data would suggest to use log-ratio uncorrelated shrinkage targets instead of the diagonal targets used for unconstrained data. We will show that it is straightforward to construct such targets. However, it is usually more convenient in practice to work with diagonal targets, for which we have to do the shrinkage on a basis of the compositions. Once the shrinkage estimate for the basis covariance is obtained, it can then be back-transformed to a logratio covariance matrix. This strategy is described in the following section.

\subsection{Basis Covariance Shrinkage Using Diagonal Targets}

A diagonal shrinkage target may often be preferred for practical reasons. As an example, the corpcor R package \cite{SchaeferStrimmer} uses the diagonal of the sample covariance as a target and does not allow for alternative targets. However, we have seen that logratio covariance matrices do not allow for this simple independence structure. Therefore, we now define the shrinkage estimator directly for the \emph{basis} covariance matrix $\boldsymbol{\Omega}$:
\begin{equation}
    \hat{\boldsymbol{\Omega}}=\lambda\boldsymbol{T}_C+(1-\lambda)\boldsymbol{C},
\end{equation}
where $\boldsymbol{C}$ is the empirical basis covariance matrix, and the shrinkage target $\boldsymbol{T}_C$ could be simply {its} diagonal. Thus, given some compositional data, our first step to construct an uncorrelated shrinkage target is to obtain an empirical basis covariance matrix from the data. For this, it is necessary to fix the size of the  basis $t$. Note that doing this specifies the variability of the totals across samples. Given that we will back-transform our estimator to a logratio covariance matrix, this variability is irrelevant. The logratio covariance structure will not contain information about the basis size, so it is convenient to use $t_i=1$ for all samples $i=1,\dots,N$, giving a basis $\boldsymbol{m}_i=\log\boldsymbol{p}_i$. In the case of relative count data, the counts themselves can provide the basis, and $\boldsymbol{C}$ can be estimated from their logarithms (provided zeros are taken care of). Yet another possibility is to start from an empirical logratio covariance matrix and obtain $\boldsymbol{C}$ via a transformation given in Table \ref{ttable}. These transformations simplify for a constant basis size.\\
 The diagonal target has the additional advantage that the expression for the optimal value of $\lambda$ simplifies \cite{SchaeferStrimmer} compared to the more general expression (\ref{lambda}). {For diagonal targets we have}
\begin{equation}
    \hat{\lambda}^*=\frac{\sum_{j=1}^D\sum_{i\ne j}\widehat{\mathrm{var}}(c_{ij})}{\sum_{j=1}^D\sum_{i\ne j}c_{ij}^2}.
\end{equation}
To obtain the shrinkage estimator of a logratio covariance matrix, all we have to do now is back-transform $\hat{\boldsymbol{\Omega}}$ to an appropriate logratio covariance matrix as specified in Table \ref{ttable}. This will also be the strategy we use for our benchmark in section \ref{benchsection}.

\subsection{Logratio Uncorrelated Targets}

While we will not use them in our benchmarks, it can be of general interest to construct LU targets for direct logratio covariance shrinkage. A simple way of constructing these targets would be as follows: pick the diagonal elements of the empirical basis covariance $\boldsymbol{C}$ as our $\alpha_i$ and then obtain the ALR target covariance by (\ref{LRuncorr}) and the CLR target covariance by (\ref{GammaLU}). However, we can also skip the construction of a basis covariance entirely and express the targets directly in terms of the logratio sample covariance matrices. For the ALR target given the empirical covariance $\boldsymbol{S}$ we find
\begin{equation}
    \tau_{ij} =
    \left\{
      \begin{array}{c@{\quad}l}
         s_{ii}-\frac{2}{D}\sum_{k=1}^{D-1}s_{ik}+\frac{2}{D^2}\sum_{k,l}s_{kl} & \mbox{if $i=j$,} \\
         \frac{2}{D^2}\sum_{k,l}s_{kl}
         & \mbox{if $i\ne j$.}
      \end{array}
    \right.\quad i,j = 1,\dots,D-1.\label{ALRtarget}
\end{equation}
Similarly, using an empirical CLR covariance matrix $\boldsymbol{G}$, the corresponding target elements $t_{ij}$ are given via the diagonal elements of $\boldsymbol{G}$ by
\begin{equation}
    t_{ij}=\left\{
      \begin{array}{c@{\quad}l}
         g_{ii}-(2g_{ii}-\frac{1}{D}\sum_kg_{kk})/D & \mbox{if $i=j$,} \\
         -(g_{ii}+g_{jj}-\frac{1}{D}\sum_kg_{kk})/D
         & \mbox{if $i\ne j$.}
      \end{array}
    \right.\qquad i,j = 1,\dots,D.\label{LUtargetCLR}
\end{equation}

\section{A Benchmark of Logratio Covariance Shrinkage}\label{benchsection}

We verify our estimators of covariance and partial correlation on both simulated data as well as single-cell gene expression data. The latter are subsampled from a ``ground truth" of greater sample size {(this strategy has also been used for microbiome data in \cite{Badri})}. Throughout, the corpcor R package from \cite{SchaeferStrimmer} is used for covariance matrix shrinkage. Note that we use the default implemented there, which consists in separately shrinking the diagonal elements of the covariance matrix with another $\lambda$ parameter, see \cite{OpgenRheinStrimmer}. The aim of the benchmark is the evaluation of our proposed procedure to first shrink a basis covariance and then transform the result to a log-ratio covariance. This is compared with a naive approach of direct log-ratio covariance shrinkage (with a diagonal target containing variances that inevitably mix parts) as well as no shrinkage at all. In the first part of this section, the logratios of our compositions are taken for granted and we just sample them from a normal distribution of rather moderate dimension ($D=40$). In this case we do not have to be concerned about zeros in the basis samples. A scenario typical for genomics data sets is treated in the second part of this section, where the basis of our compositions are high dimensional relative counts taken from single-cell gene expression data ($D=500$). There we also test various ways of dealing with zeros as well as a finite-size correction for low counts in form of frequency shrinkage.

\begin{figure}[ht]
\begin{center}
\includegraphics[width=1\textwidth]{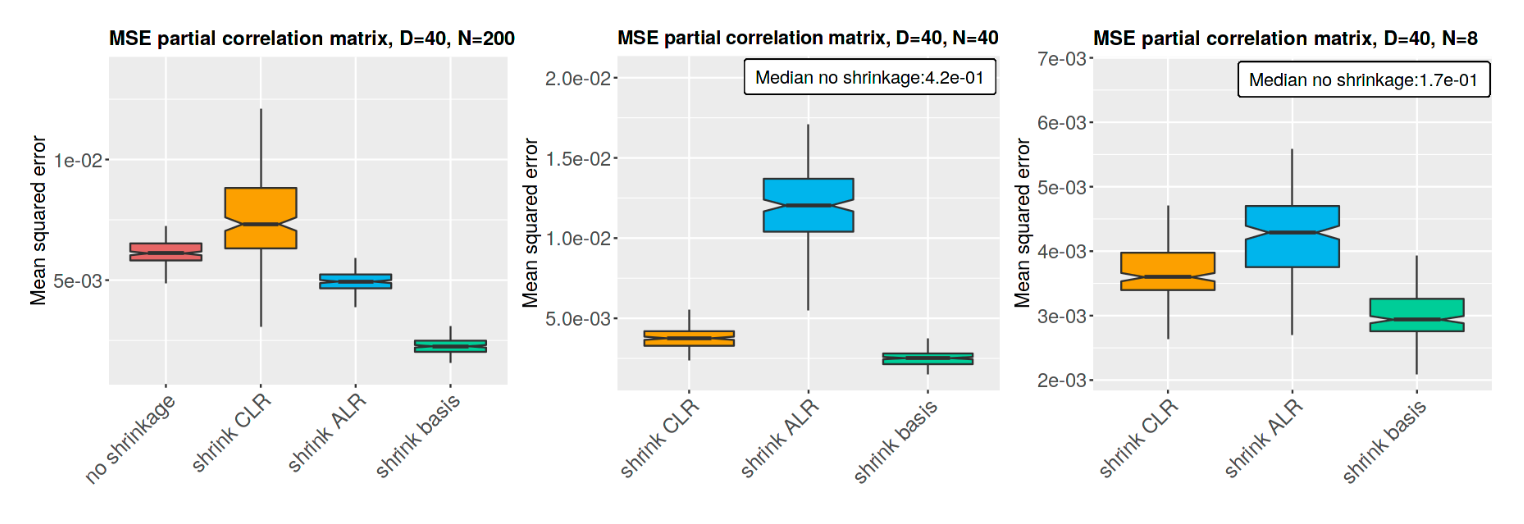}
\caption{Mean squared error of partial correlation matrix for sample sizes $N=200$, $N=40$, and $N=8$. Colours indicate different estimation procedures (no shrinkage, naive shrinkage of CLR / ALR covariance and basis {covariance} shrinkage). Each boxplot contains the results of 200 simulations. Whenever estimates without shrinkage are not shown, their median value is given in the legend.}\label{partcorr_synth}
\end{center}
\end{figure}

\subsection{Synthetic Logistic Normal Data}

\begin{figure}[H]
\centering
\includegraphics[width=1\textwidth]{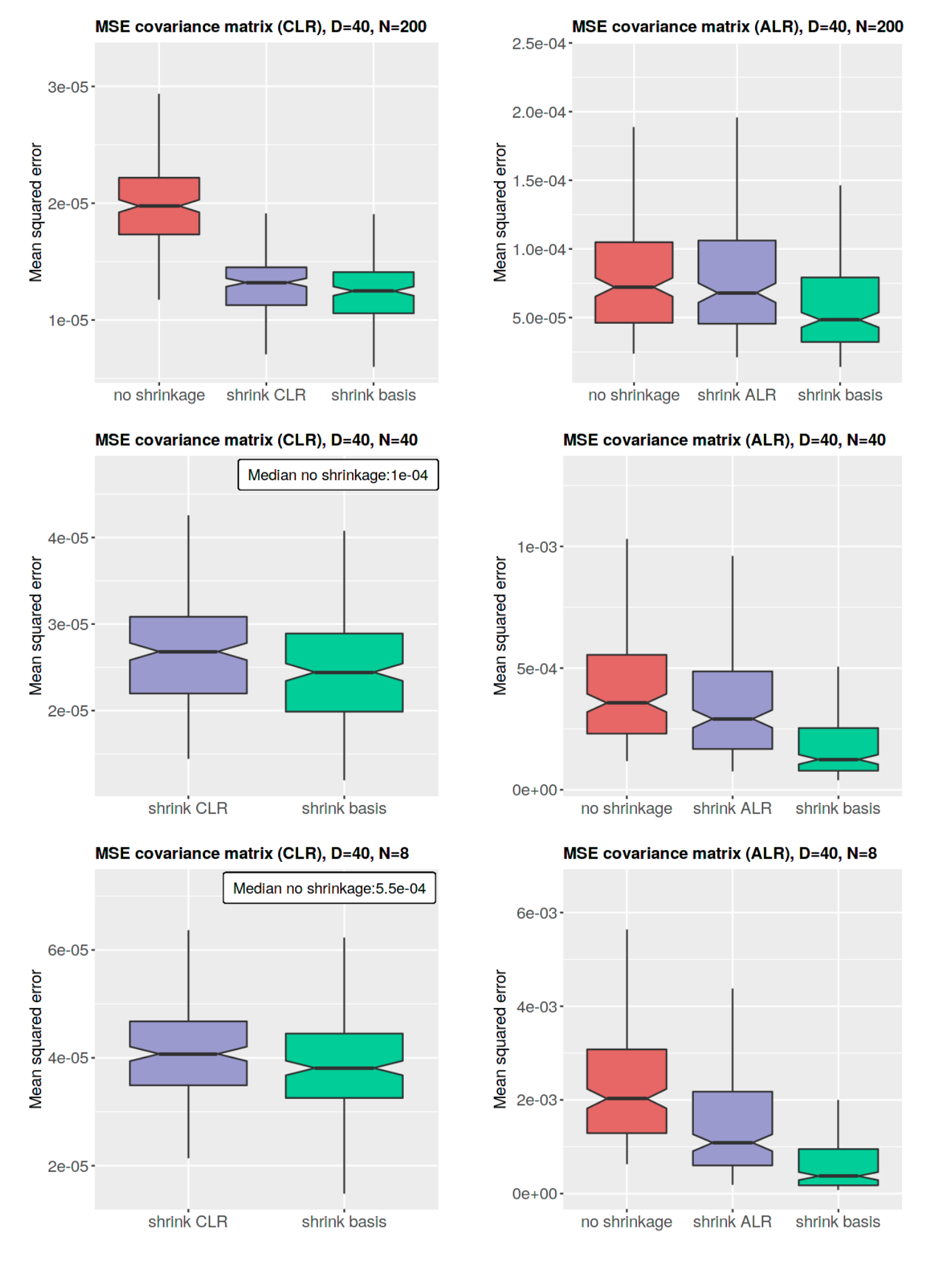}
\caption{{Mean squared error of CLR and ALR covariance matrices for different sample sizes ($N$=200, 40, 8) and estimation procedures (no shrinkage, naive shrinkage of CLR/ALR covariance matrix, and basis covariance shrinkage) computed on synthetic data.} Each boxplot contains the results of 200 simulations. Whenever estimates without shrinkage are not shown, their median value is given in the legend.}\label{cov_synth}
\end{figure}

To have reasonably realistic population parameters for the normal distribution, they are inferred from a single-cell gene expression data set \cite{Nacho}, where a subset of 240 genes that are non-zero everywhere across 5637 cells are selected. From these, in each of the 200 repetitions of the simulation, in step 1) $D=40$ parts are chosen randomly, from which ALR and CLR covariance matrices as well as the partial correlation matrix (via the pseudoinverse of the CLR covariance) are constructed across the 5K+ cells. This is our ground truth. Then 2) the ALR mean vector $\boldsymbol{\mu}$ and the ALR covariance matrix $\boldsymbol{\Sigma}$ are used to produce $N$ multivariate normal samples, where $N=8$, 40, and 200. 3) The $N$ samples are backtransformed to compositions. From these, sample estimates of CLR, ALR and (constant-size) basis covariance matrices as well as the partial correlation estimates are produced. This is done in three different ways: without shrinkage, with direct shrinkage of the logratio covariance matrices (using a diagonal target for the logratios), and via shrinkage of the basis covariance matrix (i.e., using a diagonal target for the log-transformed basis). 4) These sample estimates are compared to the ground truth established in the first step using an element-wise mean squared error. {Note that after basis covariance shrinkage, partial correlations are identical if the basis is transformed to an ALR or a CLR covariance matrix (but it is convenient to use the CLR to obtain them).}\\
The results of this procedure are seen in Figure \ref{partcorr_synth} for partial correlation and Supplementary Figure \ref{cov_synth} for covariance. There is no doubt that shrinkage leads to substantial improvements when estimating covariation, even for $N=5D$. In this regime, naive CLR shrinkage fares worse than using no shrinkage at all when estimating partial correlations. {Interestingly, the error is small for the covariance matrix itself but seems to be compounded for inference of its inverse.} In all cases we see a clear advantage when using shrinkage of the basis covariance, especially when comparing with naive ALR covariance shrinkage. We conclude that we achieve substantial improvements of our estimators when tailoring the shrinkage to the needs of logratio covariance matrices.

\begin{figure}[ht]
\begin{center}
\includegraphics[width=1\textwidth]{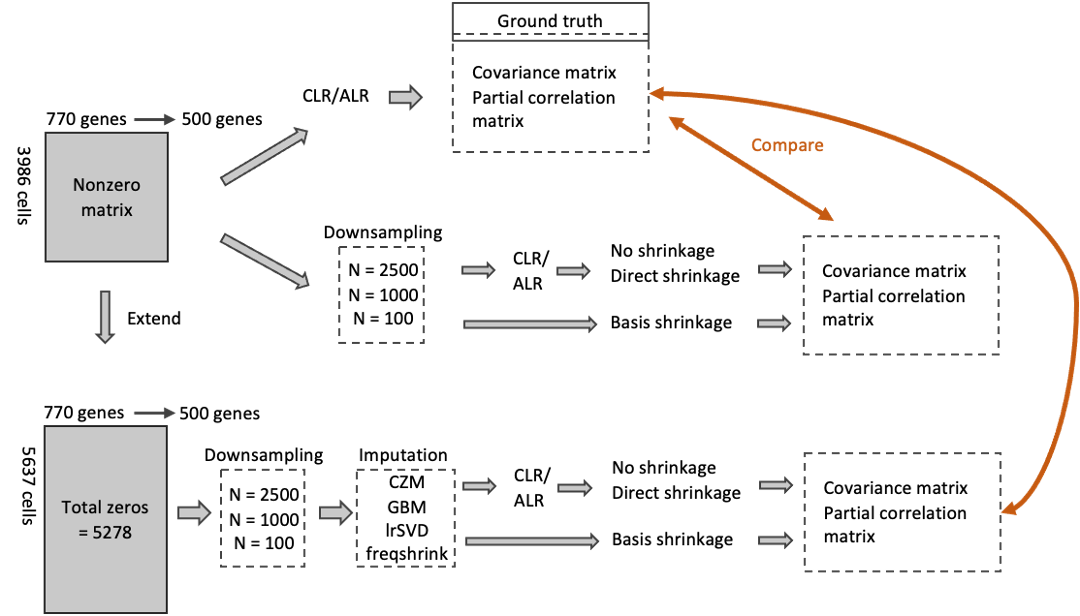}
\caption{Benchmark procedure for single-cell gene expression data.}\label{schema}
\end{center}
\end{figure}

\subsection{Single-cell gene expression data}

Instead of simulating samples from a (realistic) covariance matrix, we now draw our covariance matrices directly from the data. Since these are sample covariance matrices, the benchmark now consists in recovering them as well as possible with undersampled data. While this is not ideal (the test case could in principle be as close or closer to the population covariance than the sample covariance that serves as ground truth), it seems a reasonable assumption that the degradation effect due to undersampling is greater than the deviation of the full sample covariance from the population covariance. In the previous section we could see that shrinking the covariance still has positive effects even for $N=5D$. To avoid circularity, however, we prefer to not shrink the ground truth covariance matrix here.\\ 
\begin{figure}[ht]
\begin{center}
\includegraphics[width=1\textwidth]{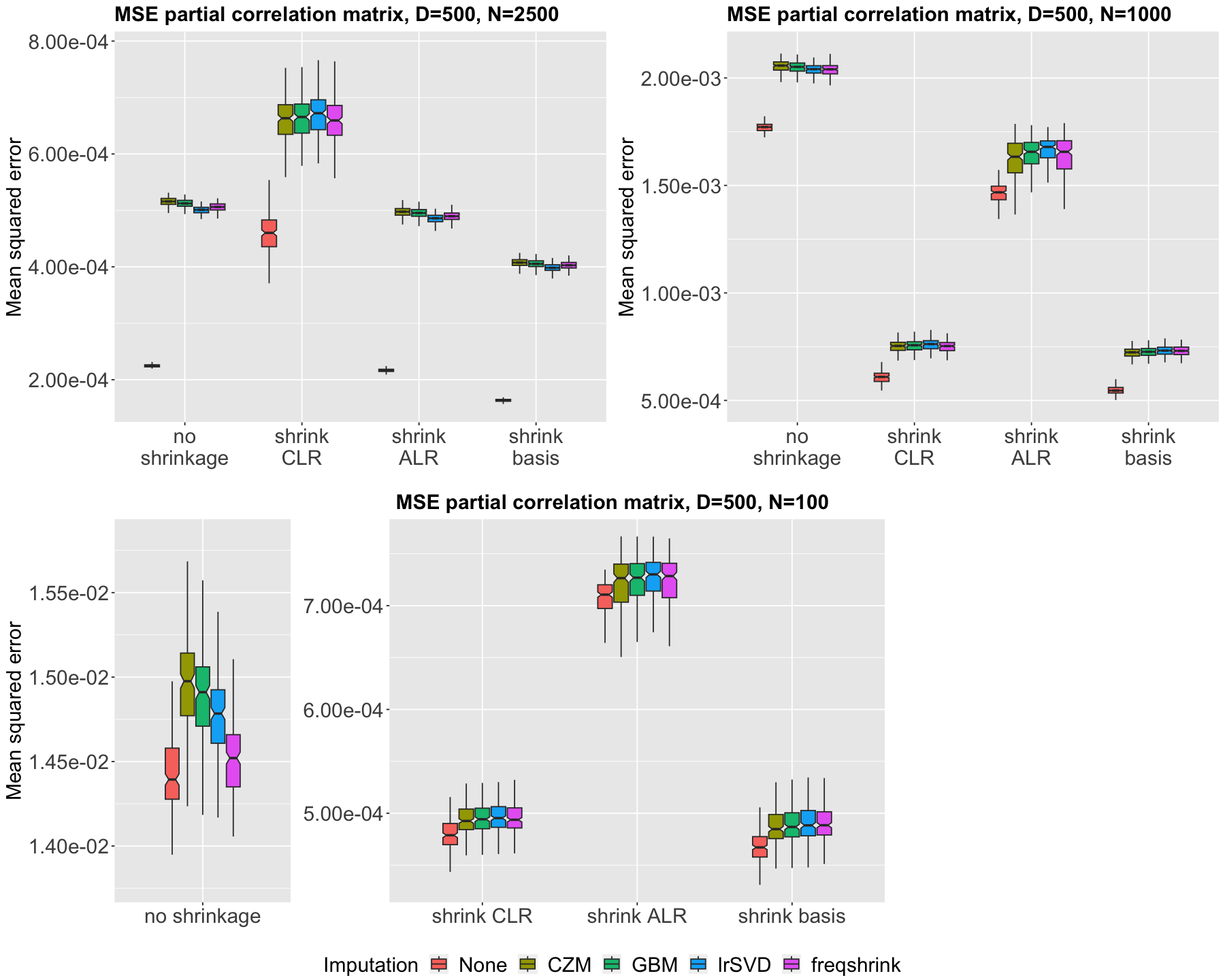}
\caption{Mean squared error of partial correlation matrices computed on {three different sizes of subsamples ($N$=2500, 1000, 100) of single-cell gene expression data. Boxplots contain 200 samples each using different estimation procedures (no shrinkage, naive ALR/CLR shrinkage, basis covariance shrinkage). Their} colour indicates the type of zero imputation used.}\label{partcorr_exp}
\end{center}
\end{figure}
Our benchmark procedure is summarized in Figure \ref{schema}. Single-cell gene expression data usually contain a very large number of zero counts, which would require strategies (e.g., pooling several cells into a single sample) that are not explored here. Instead, we concentrate on a core set of genes that are highly expressed and thus yield only a few zero counts. These can still be used to test a number of imputation strategies. We test two basic procedures implemented in the R package zCompositions \cite{zCompositions} ({Count Zero Multiplicative replacement, or CZM, and Geometric Bayesian Multiplicative replacement, or GBM}), as well as one more sophisticated strategy that takes covariance structure into account (lrSVD), which was added to zCompositions recently \cite{lrSVD}. We also test frequency shrinkage as implemented in the R package entropy \cite{HausserStrimmer}.\\
To separate the effects of shrinkage and imputation, we defined two scenarios: 1) imputation is avoided by using a large subset of cells with only nonzero counts, and 2) the effect of shrinkage and imputation are jointly evaluated by including all the cells. More precisely, we used 770 genes for which 3986 cells (out of a total of 5637 cells) have no zeros, from which a ground-truth can be constructed.
For each of the 200 resamplings, 500 genes are chosen randomly from the 770 genes. Covariance and partial correlation matrices computed on the 3986 $\times$ 500 nonzero matrix define our ground truth.
Then, in scenario 1, this matrix is subsampled to have $N$=2500, $N$=1000, and $N$=100 cells. Covariance and partial correlations computed on each of these matrices, and under the different shrinkage schemes are compared to the ground truth using mean squared error. In the second scenario, the same ground truth as defined before is used, while the subsampled data ($N$=2500, $N$=1000, and $N$=100) are drawn from the extended matrix of 5637 cells x 500 genes. Here, on average {about 1500 zeros are imputed for the 2500 samples}, 600 zeros are imputed for the 1000 samples and 60 for the 100 samples. Having additional samples that contain zeros is not ideal because these samples could in principle induce a slightly different covariance structure. The benefit, however, is that it provides a natural set of zero counts when otherwise zeros would have to be artificially introduced. Now the subsampled matrix is first imputed using four different methods, and then covariance and partial correlations are again computed under the different shrinkage schemes.\\
The results for partial correlations are shown in Figure \ref{partcorr_exp}, and for the covariance matrices in Supplementary Figure \ref{cov_exp}. Two main conclusions can be drawn from the results for such high-dimensional data. First, the effect of using the basis covariance shrinkage is strong compared with naive ALR covariance shrinkage, but it is much weaker when comparing with naive CLR covariance shrinkage. Naive shrinkage of the CLR covariance only leads to an {important} deterioration for the partial correlations {at higher sample sizes, in concordance with what we observed in the synthetic benchmark. Apart from that, the improvement is mainly observed when no zeros are imputed, which hints at a flattening of the effect caused by the imputation. The advantage of basis covariance shrinkage becomes entirely invisible when directly evaluating the covariance matrices, whose inference seems to be less prone to error than their inverse. A second main conclusion is that} imputation of zeros leads to considerable increase in mean squared error. All imputation schemes lead to a comparable loss in accuracy, with {only slight differences between  methods}.

\section{Logratio-based Partial Correlations for Wide Data}

We have addressed some questions regarding covariance structure that remained unanswered in \cite{Erb2020}, especially with respect to the application to wide data sets like the ones occurring in genomics. We think it is beneficial to also discuss some questions regarding the interpretation of logratio-based partial correlations both in CoDA and genomics. In genomics, partial correlation analysis is common, but it has rarely been based on logratio analysis \cite{Kurtz}. In CoDA, on the other hand, logratio analysis is the dominant paradigm, but partial correlations have been introduced quite recently \cite{Erb2020}. In the following sections we will discuss the relationship between classical partial correlations and their logratio counterparts.  

\subsection{Normalisations to Avoid Compositional Bias}

We can think of two strategies to avoid compositional bias when analyzing covariance structure: trying to recover the original (absolute) signal (as contained in one particular basis covariance $\boldsymbol{\Omega}$) or using only that part of the signal that absolute and relative data have in common (which is contained in the logratio covariance). The first  (``normalisation") strategy relies on assumptions because the closure operation discards a part of the signal and to recover the original totals up to a constant factor, additional information is needed. The second (``CoDA") strategy voluntarily discards this information and doing so introduces dependencies between the variables, but it is assumption-free. Interestingly, the distinction between the two approaches is not all that clear-cut because the logratio transformations Aitchison introduced can be used to pursue the ``normalisation" strategy of re-opening the data with additional assumptions. This was discussed in (Quinn et al., 2018), see especially the Supplementary material. In the following, we revisit these arguments from the perspective of logratio covariance matrices and their basis covariance $\boldsymbol{\Omega}$. Let us start with the ALR covariance $\boldsymbol{\Sigma}$. It can be obtained from the basis covariance by
\begin{equation}
    \sigma_{ij}=\omega_{ij}-\omega_{iD}-\omega_{jD}+\omega_{DD}\label{Omega2Sigma}
\end{equation}
(see Table \ref{ttable}). Note that if the reference variable $m_D$ remains unchanged across samples {\it in the basis}, the three last terms on the right-hand side vanish, and the two covariances coincide. This is the reason behind the widespread normalisation procedure in genomics where (absolutely) unchanged reference genes are used to ``normalize" the data to avoid compositional bias. Similarly, for the CLR covariance $\boldsymbol{\Gamma}$ we have the relationship 
\begin{equation}
    \omega_{ij}=\gamma_{ij}+\beta_i+\beta_j\label{Gamma2Omega}
\end{equation}
(see again Table \ref{ttable}). From the dependence of $\beta_i$ on $g(\boldsymbol{m})$ we see that this implies that if the geometric mean {\it of the basis} remains unchanged across samples, the $\beta$ terms vanish and the CLR covariance is identical with the basis covariance. This is in agreement with the notion of effective library size normalisation, where a reference that remains unchanged on the absolute data is constructed to avoid compositional bias. The geometric mean can remain unchanged across samples before closure when the majority of genes is not subject to systematic change. Under this assumption the CLR transformed data have an identical correlation structure as the log-transformed absolute data.\\
We can see that the CoDA strategy has the advantage that, while it is on the safe side if the assumptions aren't met, it also recovers the original signal if they happen to be fulfilled. In the latter case, however, a specific reference and thus logratio covariance {would be picked out because it recovers the basis covariance}. While this is necessary for evaluating both covariance and correlation, it is unnecessary for partial correlation. The results remain invariant under permuting the reference in $\boldsymbol{\Sigma}$ and they are also identical when using the pseudoinverse of $\boldsymbol{\Gamma}$ for their evaluation, see also \cite{Erb2020}.\\
These benefits of logratio-based partial correlation are lost when using more naive approaches, like deriving partial correlations from the covariance matrix of $\log\boldsymbol{p}$, or from a {\it relative} count basis using $\log\boldsymbol{m}$ directly. Formally, partial correlations derived when not specifying a reference do not coincide with their logratio counterparts.

\begin{figure}[H]
\centering
\includegraphics[width=1\textwidth]{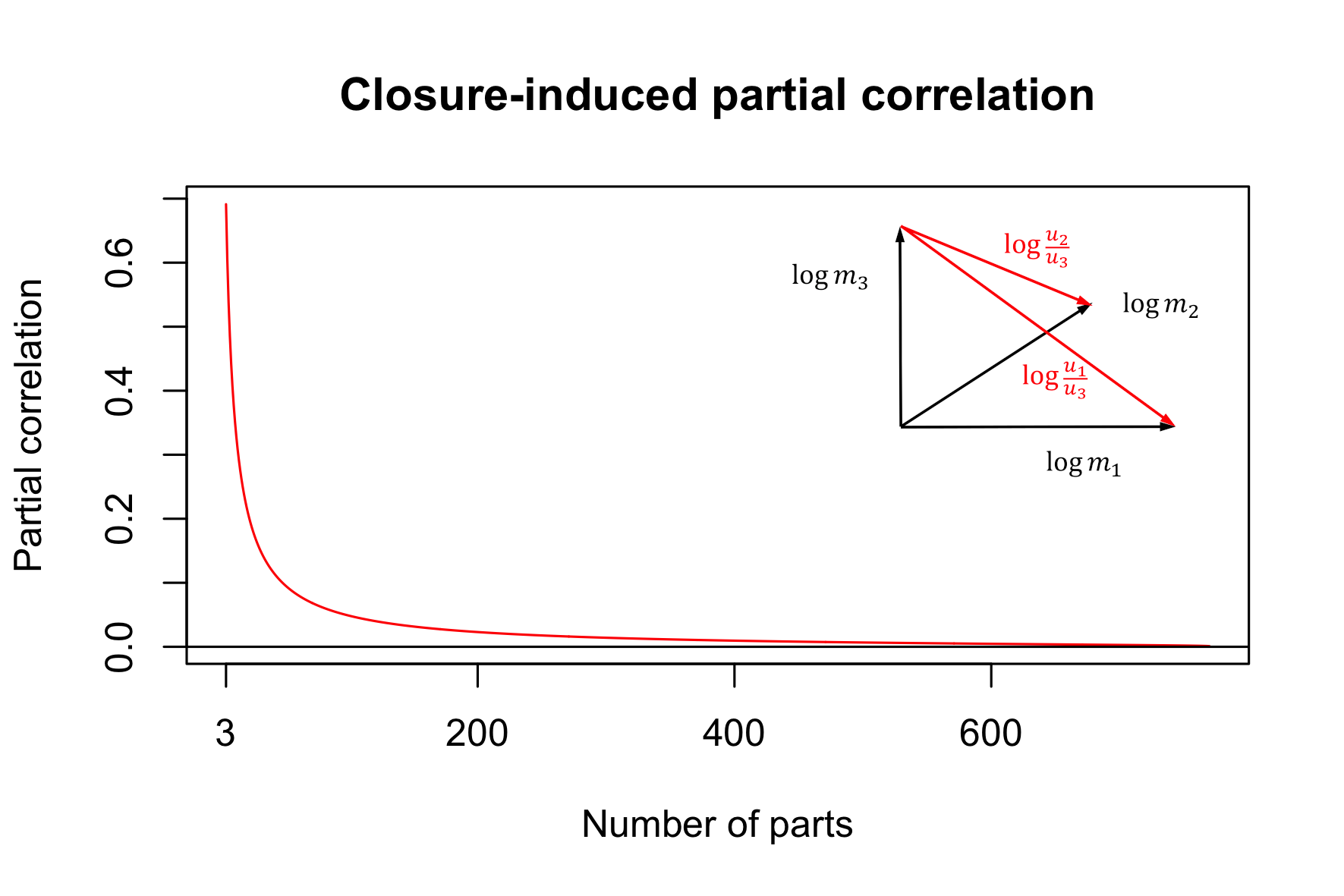}
\caption{A worst-case scenario for spurious logratio-based partial correlation of a fixed pair of parts in artificially uncorrelated data (see text). Successive increase of the number of parts from three (PC=0.69) to 770 (PC=0.001) shows how the closure-induced spurious signal tends to zero with the number of parts. Inset: The case $D=3$. The vanishing correlation between the first two parts could be recovered from the logratios with respect to the third part (indicated in red) by correlating their projections into the plane that is orthogonal to $\log m_3$. This would just be the correlation between residuals when controlling for $\log m_3$. However, the information about $\log m_3$ is lost, so partial correlation between the logratios remains biased.}\label{dilution}
\end{figure}

\subsection{What is the effect of the closure?}

Partial correlations based on logratios can emulate their absolute counterparts extremely well (a benchmark of which will be shown in another contribution). However, the closure operation makes us lose one dimension which cannot be recovered. So there remains a difference between partial correlations derived from absolute (using logarithms only) versus relative data (using logratios). Or, to put it differently, partial correlations can change when we derive them from a logratio covariance as opposed to the original (usually unknown) basis covariance. Using the relation between $\boldsymbol{\Omega}$ and $\boldsymbol{\Gamma}$ of Table \ref{ttable}, the respective covariance inverses are related by
\begin{equation}
    \boldsymbol{\Omega}^{-1}=\left(\boldsymbol{\Gamma}+\boldsymbol{B}\right)^{-1},
\end{equation}
where the $D\times D$ matrix $B$ has elements $\beta_i+\beta_j$. We see that we can neatly separate the covariance signal introduced by the variability of the basis size in the matrix $\boldsymbol{B}$, but taking the inverse will mix these signals again. To further simplify the relationship of the inverses, we need a more specific expression for the covariance. A simple way to extract the partial correlation that is only due to the closure operation is to consider a diagonal basis covariance, as it has no partial correlation by definition. We know from section \ref{LUsection} that linear independence corresponds to a {logratio uncorrelated} (LU) covariance in the embedded logratio sample space (where one dimension is lost). In other words, the closure applied to an uncorrelated basis leads to the dependence structure of LU compositions\footnote{Dirichlet compositions are LU [Aitchison 1986], and this (often undesired) property gave rise to Aitchsion's introduction of the logistic normal. The richer covariance structure of the latter enables modelling that goes beyond unit-sum induced interactions. Of course, a logistic normal with an LU covariance matrix also models LU compositions.}.\\
The elements of $\boldsymbol{\Sigma}$ for an LU composition are given by (\ref{LRuncorr}) and (\ref{alpha}). The fact that $\boldsymbol{\Sigma}$ has nonzero off-diagonal elements for LU compositions immediately shows that the partial correlations of LU compositions do not vanish. It is not difficult to determine the precise expression. For $\boldsymbol{u}$ an LU composition, its (logratio-based) partial correlations are given by
\begin{equation}
    r_{ij}(\boldsymbol{u})=\sqrt{\frac{\alpha^{-1}_i\alpha^{-1}_j}{\left(\sum_{k\ne i}\alpha^{-1}_k\right)\left(\sum_{k\ne j}\alpha^{-1}_k\right)}},\label{PCLU}
\end{equation}
for all $i\ne j$, and where $\boldsymbol{\alpha}$ is given by (\ref{alpha}) (see Appendix for a proof).\\
To get an idea how strong this ``spurious" signal is, we make the following experiment. To obtain an (artificial) system of linearly independent variables, we take the diagonal of the basis covariance matrix of the 770 genes considered in the previous section. This diagonal corresponds to $\boldsymbol{\alpha}$, and we can calculate the logratio-based partial correlations between all pairs using (\ref{PCLU}). For a worst-case scenario, we select the pair of parts for which we obtain the strongest partial correlation. With this pair fixed, we repeat calculating its partial correlation after eliminating an arbitrary part (and thus one component from the vector $\boldsymbol{\alpha}$). We repeat this process until we have only one additional part left with our original pair. The result of this process is shown in Figure \ref{dilution} (going from right to left along the horizontal axis). We see a similar ``dilution" of the compositional effect as for the negative correlations between multinomial counts, see \cite{Reappraisal}. In our case, the spurious correlation turns out to be positive, but it is evaluated between residuals of logratios, not compositional parts.

\section{{Conclusion}}

{We have proposed a simple way of obtaining shrinkage-based estimators for log-ratio covariance matrices and their inverse. As shown in our benchmark, these estimators clearly improve on the (still widely used) empirical estimators and also outperform estimators that shrink log-ratio covariance matrices towards a diagonal target as proposed in \cite{Badri}. The use of Aitchison's notion of log-ratio uncorrelated compositions is crucial to obtain the improved estimates.\\
With our contribution we also aim to promote the use of partial correlations for the analysis of compositional data sets. We see three advantages in using them: 1) They take all parts into account for the evaluation of pairwise interaction (and thus can factor out indirect interactions). 2) Their evaluation is reference-independent. 3) Negative correlations can be meaningfully evaluated with respect to the set of variables that are controlled for. While partial correlations change under taking subcompositions, this is expected because it reflects the effect of the removal of variables that are partialled out. However, we have also shown that a ``spurious" positive signal remains that is due to the loss of the $D$-th dimension in compositional data sets. We have derived the expression for this remaining partial correlation induced by the closure operation and have shown how it ``dilutes out" with a growing number of parts. In this sense, partial correlations between parts are expected to show little spurious signal whenever there are sufficiently many parts, a common occurrence in genomic data.\\
For genomic data sets, we have also drawn some connections with the problem of their normalisation. Sequencing data are relative counts, and as such their analysis within the scale-free logratio framework can be beneficial. The use of various normalisation schemes to control for sequencing depth can essentially be circumvented in this way. We see much promise in the use of logratio-based partial correlations especially for situations when normalisation assumptions fail.}   

\section{Appendix}

\subsection{Bayesian Interpretation of Covariance Shrinkage}

In the Bayesian view of statistical inference, the parameters of a distribution are considered to follow distributions themselves. The posterior distribution of the parameter in question incorporates information from both the data (via the likelihood) and the prior distribution of the parameter. Shrinkage estimators also can incorporate prior information, and while a suitable empirical estimator can maximise likelihood, optimization of shrinkage is related to maximizing the posterior of the parameter\footnote{As an example, in the case of shrinkage estimates of multinomial frequencies, the target frequencies are renormalized pseudocounts to the count data, and the shrinkage intensity $\lambda$ is the prior sample size of the conjugate Dirichlet prior that has the pseudocounts as its parameters \cite{HausserStrimmer}.}. In the following we want to make this more precise for Gaussian covariance shrinkage.\\
Let us start with the Gaussian conjugate prior distribution of the precision matrix $\boldsymbol{\Sigma}^{-1}$ in the case of a fixed mean $\boldsymbol{\mu}$. This is known as the Wishart (or multivariate Gamma) distribution with density
\begin{equation}
    f_\mathcal{W}(\boldsymbol{\Sigma}^{-1}|\boldsymbol{V},\nu)=\frac{|\boldsymbol{\Sigma}^{-1}|^{(\nu-D)/2}e^{-\frac{1}{2}\mathrm{tr}\left(\boldsymbol{\boldsymbol{V}^{-1}\Sigma}^{-1}\right)}}{2^{\nu(D-1)/2}|\boldsymbol{V}|^{\nu/2}\Gamma_{D-1}(\frac{\nu}{2})},\label{Wishart}
\end{equation}
where the number of degrees of freedom $\nu$ is a positive integer, the parametric matrix $\boldsymbol{V}$ is positive definite of order $(D-1)\times(D-1)$,  and $\Gamma$ denotes the multivariate Gamma function. The particular form of this density can be understood better when decomposing the precision matrix as
\begin{equation}
    \boldsymbol{\Sigma}^{-1}=\boldsymbol{U}\boldsymbol{U}^T,
\end{equation}
where $\boldsymbol{U}$ is of order $(D-1)\times\nu$. If the $\nu$ vectors in $\boldsymbol{U}$ are independently drawn from a normal distribution with mean zero and covariance $\boldsymbol{V}$ (where $\nu\ge D-1$), then 
the precision matrix follows a Wishart distribution with density (\ref{Wishart}).\\
To estimate a Gaussian covariance matrix, we need more than a single sample of data. Let us now consider a data matrix $\boldsymbol{X}$ where the $N$ rows were sampled according to (\ref{Gauss}). The joint likelihood of these samples can be written in terms of their sample parameter estimates $\boldsymbol{S}$ and $\bar{\boldsymbol{x}}$ as
\begin{multline}
    f_\mathcal{N}(\boldsymbol{X}|\boldsymbol{\mu},\boldsymbol{\Sigma})=\prod_{i=1}^Nf_\mathcal{N}(\boldsymbol{x}_i|\boldsymbol{\mu},\boldsymbol{\Sigma})=\\
    \left(\frac{(2\pi)^{D-1}}{|\boldsymbol{\Sigma}|}\right)^\frac{N}{2}\mathrm{exp}\left\{-\frac{1}{2}\mathrm{tr}\left[\boldsymbol{\Sigma}^{-1}\left((N-1)\boldsymbol{S}+N(\bar{\boldsymbol{x}}-\boldsymbol{\mu})^T(\bar{\boldsymbol{x}}-\boldsymbol{\mu})\right)\right]\right\}.
\end{multline}
Let us now assign the conjugate priors as follows:
\begin{eqnarray}
    \boldsymbol{\mu}~|~\boldsymbol{\Sigma}^{-1}&\sim&\mathcal{N}\left(\boldsymbol{\mu}_0,\frac{1}{\kappa}\boldsymbol{\Sigma}\right),\\
    \boldsymbol{\Sigma}^{-1}&\sim&\mathcal{W}(\boldsymbol{V},\nu).
\end{eqnarray}
The joint posterior density then factorizes as $P(\boldsymbol{\mu}| \boldsymbol{\Sigma}^{-1})P(\boldsymbol{\Sigma}^{-1})$, where the marginal $P(\boldsymbol{\Sigma}^{-1})$ is again Wishart. More precisely
\begin{equation}
    P(\boldsymbol{\Sigma}^{-1}|\boldsymbol{S},\bar{\boldsymbol{x}},\boldsymbol{\mu}_0,\kappa,\boldsymbol{V},\nu)=f_\mathcal{W}(\boldsymbol{\Sigma}^{-1}|\boldsymbol{V}^*,\nu+N),
\end{equation}
where
\begin{equation}
    \boldsymbol{V}^*=(N-1)\boldsymbol{S}+\boldsymbol{V}+\frac{\kappa N}{\kappa+N}(\bar{\boldsymbol{x}}-\boldsymbol{\mu}_0)^T(\bar{\boldsymbol{x}}-\boldsymbol{\mu}_0)\label{postSig}
\end{equation}
(see \cite{deGroot}, p.178)\footnote{For the sake of completeness, it should perhaps be mentioned that $P(\boldsymbol{\mu}| \boldsymbol{\Sigma}^{-1})=\\
f_\mathcal{N}\left((\kappa\boldsymbol{\mu}_0+N\bar{\boldsymbol{x}})/(\kappa+N),\boldsymbol{\Sigma}/(\kappa+N)\right)$.}. It follows that under the Bayesian model, an improved covariance estimator will involve an additive correction to the sample covariance. We can now write (\ref{postSig}) in terms of the shrinkage estimator when taking into account the posterior sample size:
\begin{equation}
    (\nu+N)\hat{\boldsymbol{\Sigma}}=\boldsymbol{V}^*.\label{BayesShrink}
\end{equation}
Identifying (\ref{BayesShrink}) with (\ref{shrink}), we obtain the identities
\begin{eqnarray}
    (1-\lambda)&=&(N-1)/(\nu+N),\\
    \lambda\boldsymbol{T}&=&\frac{1}{\nu+N}\left(\boldsymbol{V}+\frac{\kappa N}{\kappa+N}(\bar{\boldsymbol{x}}-\boldsymbol{\mu}_0)^T(\bar{\boldsymbol{x}}-\boldsymbol{\mu}_0)\right),
\end{eqnarray}
from which it follows that
\begin{eqnarray}
    \lambda&=&\frac{\nu+1}{\nu+N},\\
    \boldsymbol{T}&=&\frac{1}{\nu+1}\left(\boldsymbol{V}+\frac{\kappa N}{\kappa+N}(\bar{\boldsymbol{x}}-\boldsymbol{\mu}_0)^T(\bar{\boldsymbol{x}}-\boldsymbol{\mu}_0)\right).
\end{eqnarray}
Optimization of $\lambda$ thus boils down to a tuning of the prior degrees of freedom $\nu$ relative to the sample size $N$, while the target $\boldsymbol{T}$ defines a prior covariance structure.

\subsection{Proof of (\ref{LRuncorr}) for LU compositions}
We start with the case $i\ne j$ (see [Aitchison 1986], section 5.9):
\begin{multline}
    \sigma_{ij}=\mathrm{cov}\left(\log\frac{p_i}{p_D},\log\frac{p_j}{p_D}\right)=\mathrm{cov}\left(\log\frac{p_i}{p_k}+\log\frac{p_k}{p_D},\log\frac{p_j}{p_D}\right)\\\
    =0+\sigma_{kj}
\end{multline}
because of (\ref{LUcondition}). It is therefore clear that $\sigma_{ij}$ is constant for unequal indices and the definition (\ref{alpha}) makes sense for LU compositions. So $\sigma_{ij}=\alpha_D$ if $i\ne j$. Similarly, in the case $i=j$ we have
\begin{multline}
    \sigma_{ii}=\mathrm{var}\left(\log\frac{p_i}{p_D}\right)\\
    =\mathrm{cov}\left(\log\frac{p_i}{p_D},\log\frac{p_i}{p_D}\right)=\mathrm{cov}\left(\log\frac{p_i}{p_k}+\log\frac{p_k}{p_D},\log\frac{p_i}{p_D}\right)\\
    =\mathrm{cov}\left(\log\frac{p_k}{p_i},\log\frac{p_D}{p_i}\right)+\mathrm{cov}\left(\log\frac{p_k}{p_D},\log\frac{p_i}{p_D}\right)=\alpha_{i}+\alpha_D,
\end{multline}
which concludes the proof.

\subsection{Proof of (\ref{alphaOmega}) for an Uncorrelated Basis}
Let the composition $\boldsymbol{u}$ have a basis $t\boldsymbol{u}$ with $\mathrm{cov}(\log (tu_i),\log (tu_j))=0$ for all $i\ne j$. We show that it follows that
\begin{equation}
    \mathrm{var}\left(\log (t u_k)\right)=\mathrm{cov}\left(\log\frac{u_i}{u_k},\log\frac{u_j}{u_k}\right)
\end{equation}
where none of the indices $i$, $j$, $k$ are equal. We write the left-hand side as
\begin{multline}
    \mathrm{cov}\left(\log (t u_k),\log (t u_k)\right)=\mathrm{cov}\left(\log\frac{tu_k}{tu_i}+\log(tu_i),\log\frac{tu_k}{tu_j}+\log (t u_j)\right)\\
    =\mathrm{cov}\left(\log\frac{tu_k}{tu_i},\log\frac{tu_k}{tu_j}\right)+\mathrm{cov}\left(\log\frac{tu_k}{tu_i},\log(tu_j)\right)\\
    +\mathrm{cov}\left(\log(tu_i),\log\frac{tu_k}{tu_j}\right)+\mathrm{cov}\left(\log(tu_i),\log(tu_j)\right)\\
    =\mathrm{cov}\left(\log\frac{u_i}{u_k},\log\frac{u_j}{u_k}\right)+\mathrm{cov}\left(\log(tu_k)-\log(tu_i),\log(tu_j)\right)+\dots
\end{multline}
The terms after the first one are all zero because of the condition that $\mathrm{cov}(\log(tu_i),\log(tu_j))=0$ for any indices $i\ne j$, which concludes the proof.

\subsection{Proof of (\ref{PCLU}) for LU compositions}

Let $\boldsymbol{\Sigma}$ be given by the elements (\ref{LRuncorr}). The following are expressions for its determinant (here given without proof):
\begin{equation}
    |\boldsymbol{\Sigma}|=\prod_{i=1}^D\alpha_i\sum_{i=1}^D\frac{1}{\alpha_i}=\sum_{i=1}^D\prod_{k\ne i}\alpha_k,\label{detSigma}
\end{equation}
where the second equality is found by evaluating the greatest common denominator of the $\alpha_i^{-1}$. We show that the matrix inverse of $\boldsymbol{\Sigma}$ is given by the elements 
\begin{equation}
    \sigma_{ij}^{(-1)} =\frac{1}{\sum_{k=1}^D\prod_{l\ne k}\alpha_l}
    \left\{
      \begin{array}{c@{\quad}l}
        \sum_{k\ne i}\prod_{l\ne k,i}\alpha_l  & \mbox{if $i=j$,} \\
         -\prod_{k\ne i,j}\alpha_k
         & \mbox{if $i\ne j$.}
      \end{array}
    \right.\quad i,j = 1,\dots,D-1.\label{inverseSigma}
\end{equation}
That this is the inverse can be easily proven when multiplying the resulting matrix with the matrix whose elements are given by (\ref{LRuncorr}). The off-diagonal elements (where $i\ne j$) of the result are
\begin{multline}
    \sum_{m=1}^{D-1}\sigma_{im}\sigma^{(-1)}_{mj}=\sum_{m\ne i,j,D}\sigma_{im}\sigma_{mj}^{(-1)}+\sigma_{ii}\sigma_{ij}^{(-1)}+\sigma_{ij}\sigma_{jj}^{(-1)}=\\
    \frac{1}{\sum_{k=1}^D\prod_{l\ne k}\alpha_l}\left[-\alpha_D\sum_{m\ne i,j,D}\prod_{k\ne m,j}\alpha_k-(\alpha_i+\alpha_D)\prod_{k\ne i,j}\alpha_k+\alpha_D\sum_{k\ne j}\prod_{l\ne k,j}\alpha_l\right]\nonumber
\end{multline}
The square bracket evaluates to
\begin{multline}
-\alpha_D\sum_{m\ne j,D}\prod_{k\ne m,j}\alpha_k-\alpha_i\prod_{k\ne i,j}\alpha_k+\alpha_D\sum_{k\ne j}\prod_{l\ne k,j}\alpha_l\\
=-\alpha_D\sum_{m\ne j,D}\prod_{k\ne m,j}\alpha_k-\alpha_i\prod_{k\ne i,j}\alpha_k+\alpha_D\sum_{k\ne j,D}\prod_{l\ne k,j}\alpha_l+\alpha_D\prod_{l\ne D,j}\alpha_l=0.
\end{multline}
Thus the off-diagonal elements vanish. For the diagonal elements we have
\begin{multline}
    \sum_{m=1}^{D-1}\sigma_{im}\sigma^{(-1)}_{mi}=\sum_{m\ne i,D}\sigma_{im}\sigma_{mi}^{(-1)}+\sigma_{ii}\sigma_{ii}^{(-1)}=\\
    \frac{1}{\sum_{k=1}^D\prod_{l\ne k}\alpha_l}\left[-\alpha_D\sum_{m\ne i,D}\prod_{k\ne m,i}\alpha_k+(\alpha_i+\alpha_D)\sum_{k\ne i}\prod_{l\ne k,i}\alpha_l\right].\label{diagonal}
\end{multline}
The square bracket evaluates to
\begin{multline}
    -\alpha_D\sum_{m\ne i,D}\prod_{k\ne m,i}\alpha_k+\alpha_i\sum_{k\ne i}\prod_{l\ne k,i}\alpha_l+\alpha_D\sum_{k\ne i,D}\prod_{l\ne k,i}\alpha_l+\alpha_D\prod_{l\ne D,i}\alpha_l\\
    =\alpha_i\sum_{k\ne i}\prod_{l\ne k,i}\alpha_l+\alpha_D\prod_{l\ne D,i}\alpha_l=\sum_{k\ne i}\prod_{l\ne k}\alpha_l+\prod_{l\ne i}\alpha_l=\sum_k\prod_{l\ne k}\alpha_k.\nonumber
\end{multline}
Inserting this back into (\ref{diagonal}) shows that the diagonal elements are 1, proving the inverse. We can now evaluate the partial correlation coefficient (\ref{partcorr}) for an LU composition $\boldsymbol{u}$ by inserting the expression for the inverse (\ref{inverseSigma}) and (to obtain the third equality below) making use of the second equality in (\ref{detSigma}):
\begin{multline}
   r_{ij}(\boldsymbol{u})=\frac{-\sigma^{(-1)}_{ij}}{\sqrt{\sigma^{(-1)}_{ii}\sigma^{(-1)}_{jj}}}=\frac{\prod_{k\ne i,j}\alpha_k}{\sqrt{\left(\sum_{k\ne i}\prod_{l\ne k,i}\alpha_l\right)\left(\sum_{k\ne j}\prod_{l\ne k,j}\alpha_l\right)}}\\
   =\frac{\prod_{k\ne i,j}\alpha_k}{\sqrt{\left(\prod_{l\ne i}\alpha_l\sum_{k\ne i}\alpha_l^{-1}\right)\left(\prod_{l\ne j}\alpha_l\sum_{k\ne j}\alpha_l^{-1}\right)}}\\
   =\frac{\prod_{k\ne i,j}\alpha_k}{\left(\prod_{k\ne i,j}\alpha_k\right)\sqrt{\left(\alpha_j\sum_{k\ne i}\alpha_l^{-1}\right)\left(\alpha_i\sum_{k\ne j}\alpha_l^{-1}\right)}}\\
   =\sqrt{\frac{\alpha^{-1}_i\alpha^{-1}_j}{\left(\sum_{k\ne i}\alpha^{-1}_k\right)\left(\sum_{k\ne j}\alpha^{-1}_k\right)}},
\end{multline}
which is the expression given in (\ref{PCLU}).

\subsection{Supplementary Figures}

\begin{figure}[H]
\centering
\includegraphics[width=1\textwidth]{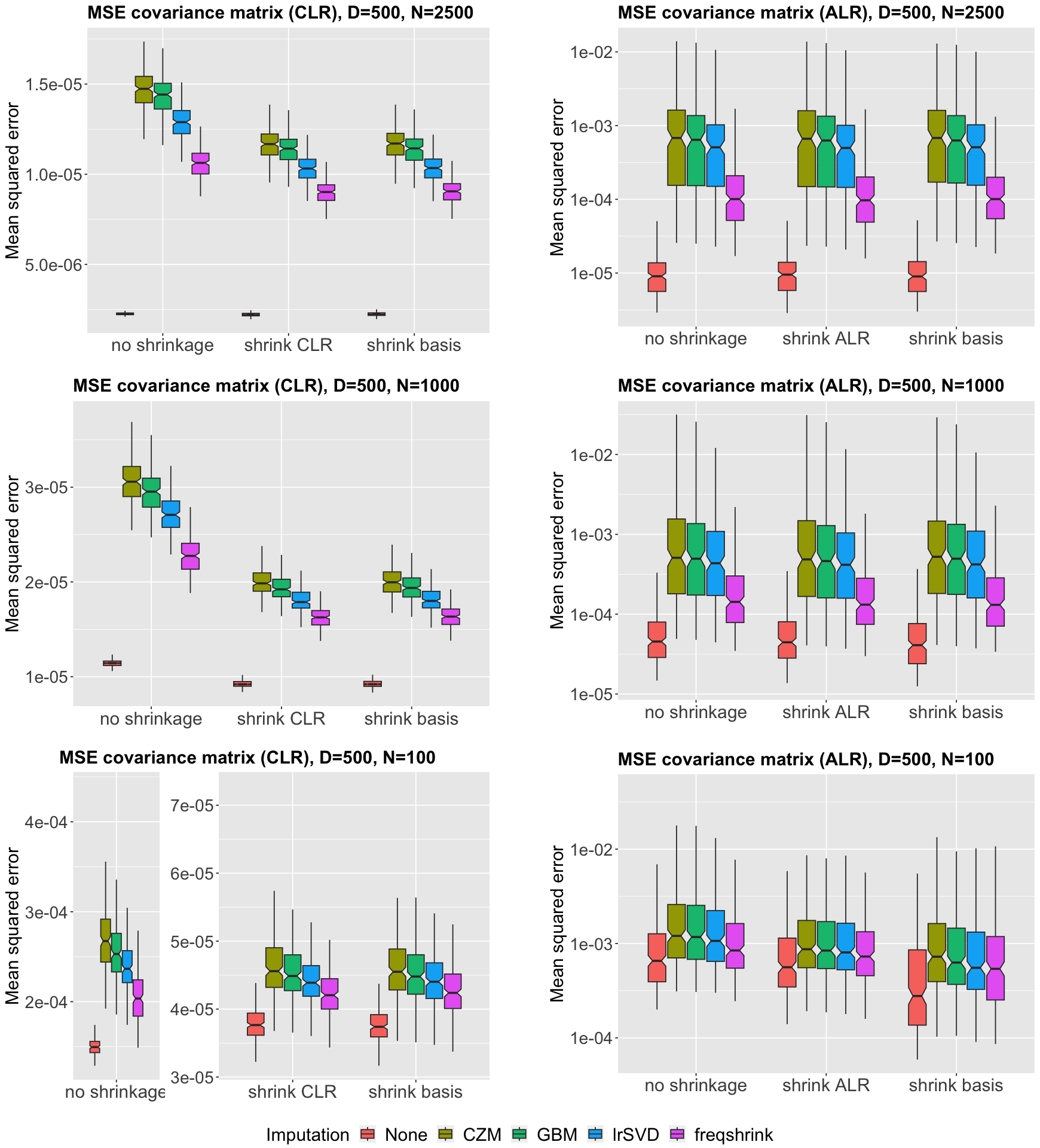}
\caption{{Mean squared error of CLR and ALR covariance matrices for different sample sizes ($N$=2500, 1000, 100) and estimation procedures (no shrinkage, naive shrinkage of CLR/ALR covariance matrix, and basis covariance shrinkage) computed on single-cell gene expression data. Each boxplot contains the results of 200 resamplings from data. Colours indicate the type of zero imputation used.}}\label{cov_exp}
\end{figure}

\subsection{Code and Data Availability}

All the code used to perform the benchmark and reproduce the results of this paper, as well as the subset of single-cell gene expression data, are available on GitHub under {https://github.com/suzannejin/pcor.bshrink.git}.\\
The  R package propr \cite{propr} enables compositional data analysis on relative gene expression data. Originally designed for efficient calculations of proportionality indices, it has been updated several times, e.g., to include differential proportionality across groups \cite{diffprop}. {Until the full implementation of partial correlations in propr, an R implementation for the computation of partial correlations with basis shrinkage is available on the repository above (bShrink function in file bin/rlib/shrink.R).}

\bibliographystyle{chicago}

\end{document}